\documentclass{aa}
\usepackage[dvipsnames,svgnames,x11names]{xcolor}

\usepackage{graphicx}
\usepackage{txfonts}
\usepackage{natbib}
\usepackage[colorlinks=true, linktocpage, linkcolor={blue!60!black}, citecolor={blue!60!black}, urlcolor={blue!60!black}]{hyperref}

\usepackage{siunitx}
\usepackage{bm}

\newcommand*{\ra}[2][]{%
    \ang[
        angle-symbol-over-decimal,
        math-degree=\textsuperscript{h},
        text-degree=\textsuperscript{h},
        math-arcminute=\textsuperscript{m},
        text-arcminute=\textsuperscript{m},
        math-arcsecond=\textsuperscript{s},
        text-arcsecond=\textsuperscript{s},
        #1]{#2}%
}

\newcommand*{\dec}[2][]{%
    \ang[angle-symbol-over-decimal,#1]{#2}%
}

\begin{document} 

\title{An ALMA+ACA measurement of the shock in the Bullet Cluster}

\author{Luca Di Mascolo\inst{1}
      \and 
      Tony Mroczkowski\inst{2} 
      \and 
      Eugene Churazov\inst{1,3} 
      \and 
      Maxim Markevitch\inst{4} 
      \and
      Kaustuv Basu\inst{5}
      \and 
      Tracy~E. Clarke\inst{6}
      \and 
      Mark Devlin\inst{7}
      \and 
      Brian~S. Mason\inst{8}
      \and 
      Scott~W. Randall\inst{9}
      \and 
      Erik~D. Reese\inst{10}
      \and 
      Rashid Sunyaev\inst{1,3}
      \and 
      Daniel~R. Wik\inst{11}
      }
      
\authorrunning{L. Di Mascolo et al.}

\institute{Max-Planck-Institut f\"{u}r Astrophysik (MPA), Karl-Schwarzschild-Strasse 1, Garching 85741, Germany\\\email{lucadim@mpa-garching.mpg.de}
     \and
         European Southern Observatory (ESO), Karl-Schwarzschild-Strasse 2, Garching 85748, Germany
     \and
         Space Research Institute, Profsoyuznaya 84/32, Moscow 117997, Russia
     \and 
         Astrophysics Science Division, NASA Goddard Space Flight Center, Greenbelt, MD 20771, USA
     \and 
         Argelander Institut f\"ur Astronomie, Universit\"at Bonn, Auf dem H\"ugel 71, D-53121 Bonn, Germany
     \and
         Naval Research Laboratory, 4555 Overlook Avenue SW, Code 7213, Washington, DC 20375, USA
     \and
         Department of Physics and Astronomy, University of Pennsylvania, 209 South 33rd Street, Philadelphia, PA 19104, USA
     \and
         National Radio Astronomy Observatory (NRAO), 520 Edgemont Road, Charlottesville, VA 22903, USA
     \and
         Harvard-Smithsonian Center for Astrophysics, 60 Garden Street, Cambridge, MA 02138, USA
     \and
         Moorpark College, 7075 Campus Rd., Moorpark, CA 93021, USA
     \and
         Department of Physics and Astronomy, The University of Utah, Salt Lake City, Utah, USA
         }

\date{Received 26 June 2019; accepted 19 July 2019}

\abstract
    {The thermal Sunyaev-Zeldovich (SZ) effect presents a relatively new tool for characterizing galaxy cluster merger shocks, traditionally studied through X-ray observations. Widely regarded as the ``textbook example'' of a cluster merger bow shock, the western, most-prominent shock front in the Bullet Cluster (1E0657-56) represents the ideal test case for such an SZ study.}
    {We aim to characterize the shock properties using deep, high-resolution interferometric SZ effect observations in combination with priors from an independent X-ray analysis.}
    {Our analysis technique relies on the reconstruction of a parametric model for the SZ signal by directly and jointly fitting data from the Atacama Large Millimeter/submillimeter Array (ALMA) and Atacama Compact Array (ACA) in Fourier space.}
    {The ALMA+ACA data are primarily sensitive to the electron pressure difference across the shock front. To estimate the shock Mach number $\mathcal{M}$, this difference can be combined with the value for the upstream electron pressure derived from an independent \emph{Chandra} X-ray analysis. In the case of instantaneous electron-ion temperature equilibration, we find $\mathcal{M}=2.08\substack{+0.12\\{-0.12}}$, in $\sim 2.4\sigma$ tension with the independent constraint from \emph{Chandra}, $\mathcal{M}_{\mathrm{X}}=2.74\pm0.25$. The assumption of purely adiabatic electron temperature change across the shock leads to $\mathcal{M}=2.53\substack{+0.33\\-0.25}$, in better agreement with the X-ray estimate $\mathcal{M}_{\mathrm{X}}=2.57\pm0.23$ derived for the same heating scenario.}
    {We have demonstrated that interferometric observations of the thermal SZ effect provide constraints on the properties of the shock in the Bullet Cluster that are highly complementary to X-ray observations. The combination of X-ray and SZ data yields a powerful probe of the shock properties, capable of measuring $\mathcal{M}$ and addressing the question of electron-ion equilibration in cluster shocks. Our analysis is however limited by systematics related to the overall cluster geometry and the complexity of the post-shock gas distribution. To overcome these limitations, a simultaneous, joint-likelihood analysis of SZ and X-ray data is needed.}
        

\maketitle

\section{Introduction}

Mergers play a crucial role in the formation of galaxy clusters, which are situated at intersections of the Cosmic Web. These spectacular events can have a profound impact on the intracluster medium (ICM) and the galaxies within these environments \citep[see, e.g.,][]{Kravtsov2012}. Mergers provide large-scale astrophysical laboratories for plasmas where the mean free path can be substantial \citep[see, e.g.,][]{Markevitch2007} and for measuring the self-interaction cross-section of dark matter \citep{Markevitch2004,Randall2008,Wittman2018,Tulin2018}. Furthermore, the very existence of dark matter was conclusively demonstrated through the merging cluster 1E0657--56,  or ``Bullet Cluster'', which exhibits spatial offsets between its baryonic and total mass peaks in the X-ray and gravitational lensing maps \citep{Clowe2006,Bradac2006,Paraficz2016}.

Key to identifying merging clusters is the detection of shocks in the ICM. A ``textbook example of a bow shock'' is observed in the X-ray image of the Bullet Cluster \citep{Markevitch2002}. Using 500~ks of \emph{Chandra} X-ray data, \cite{Markevitch2006} reported a Mach number ${\cal M}=3.0\pm0.4$ for the western, most-prominent shock in the Bullet Cluster, an estimate largely determined by the density jump conditions. We also note that \cite{Shimwell2015} revealed a second shock, on the eastern (opposite) side of the cluster, which we do not consider here.

Here we present deep, continuum Atacama Large Millimeter/submillimeter Array (ALMA) observations, sensitive to the thermal Sunyaev-Zeldovich \citep[tSZ;][]{Sunyaev1972} effect, of the main shock in the Bullet Cluster. These observations include data from both the 12-meter array (hereafter ``ALMA'') and 7-meter Atacama Compact (Morita) Array (ACA). As the tSZ effect is linearly sensitive to the line-of-sight integral of the electron thermal pressure 
(see, e.g., \mbox{\citealt{Carlstrom2002}} or \citealt{Mroczkowski2019} for reviews), these observations complement the X-ray constraints on plasma density and, less accurately, electron temperature, yielding a ground-based, mm-wave view of the shock properties.

All the results presented in this work have been derived assuming a flat $\Lambda$CDM cosmology with $\Omega_m=0.30$, $\Omega_{\Lambda}=0.70$, and $H_0=70.0~\mathrm{km~s^{-1}~Mpc^{-1}}$. At the redshift of the Bullet Cluster ($z = 0.296$), $1\arcsec$ corresponds to a physical scale of $4.41~\mathrm{kpc}$. Unless stated differently, all reported best-fitting parameters and their respective uncertainties are obtained from the $50^{\mathrm{th}}$, $16^{\mathrm{th}}$, and $84^{\mathrm{th}}$ percentiles of the marginalized posterior distributions, corresponding to the 68\% credibility interval of the distribution.\footnote{In the case of Gaussian uncertainties, the $50^{\mathrm{th}}$ percentile corresponds to the median value, and the $16^{\mathrm{th}}$ and $84^{\mathrm{th}}$ percentiles correspond to $-1\sigma$ and $+1\sigma$ deviations from this.}

\section{Data and analysis overview}
\subsection{ALMA/ACA observations}
As part of ALMA Cycle 2 operations ALMA and the ACA observed the Bullet Cluster for a total of 3.1 and 5.9~hours integration time respectively in Band~3 (project code 2013.1.00760.S). These wideband observations span the frequency range $84-100~\mathrm{GHz}$ in four $2~\mathrm{GHz}$-wide spectral windows, centered at approximately $85$, $87$, $97$, and $99~\mathrm{GHz}$. Our strategy employed a single, deep observation centered approximately on the nose of the shock front, as inferred from the X-ray observations \citep{Clowe2006,Markevitch2006}. 
The ALMA and ACA observations were respectively performed in 4 and 11 separate executions spanning 2014, obtaining root-mean-square noise levels of approximately $5~\mu\mathrm{Jy}$ and $45~\mu\mathrm{Jy}$ respectively (as measured in naturally-weighted imaging), and a synthesized beam with a main lobe of $4.01\arcsec\times3.07\arcsec$ FWHM (P.A.\ $81\degr$).

We re-reduced the data using the ALMA pipeline \citep{ALMA2015,ALMA2016} in CASA~4.7 \citep{McMullin2007}, producing results consistent with the previous calibration using the script provided on data delivery. Our re-reduction provided a cross-check of the earlier reduction, and was necessary due to backwards-compatibility issues and bug fixes in subsequent CASA releases. The data were calibrated using the default calibration strategy of the ALMA observatory, which has nominal uncertainty $\leq 5\%$. However, since the flux calibrators, which included quasars, differed for each execution, we performed a manual cross-check of the values for the designated flux calibrators as well as the phase calibrators, finding they were consistent for the dates spanned by the observations.

An interferometer behaves as a spatial filter, sampling the sky only in Fourier modes corresponding to projected baselines in the array \citep[see, e.g.,][]{Condon2016,DiMascolo2019}. This provides clean imaging free from atmospheric structure, but also leads to two major complexities: incomplete sampling in Fourier space even for the modes accessible to the array, and the lack of recovery of angular scales larger than those corresponding to the shortest projected distances between array elements (i.e., the ``missing flux issue''). Based on the $uv$-space coverage of the ALMA and ACA data presented here, the largest recoverable scales are respectively $\sim40\arcsec$ and $\sim55\arcsec$. As detailed in the next section, we choose to forward-model the observed SZ signal using X-ray-motivated priors to address such issues. Additionally, to avoid known deconvolution biases intrinsic to the \textsc{clean} algorithm \citep{Hogbom1974, Thompson2015}, we perform our analysis directly in visibility ($uv$) space. We extend the interferometric SZ analysis techniques presented in the Appendix of \citet{DiMascolo2019} to allow fitting pressure discontinuities due to shocks. Our approach builds upon the work of \citet{Basu2016}, but incorporates several advances in the parameter-space sampling technique as well as more sophisticated and flexible models allowed by the deeper X-ray and SZ observations. In brief, we build an image-space model of the SZ signal by integrating numerically the three-dimensional pressure distribution model, and applying the proper SZ frequency scaling. The dependence of the SZ signal on the electron temperature is taken into account when modelling the SZ spectrum \citep[][and end of Sec.~\ref{sec:gnfw}]{Itoh2004,Chluba2012}. The pixel scale is chosen to fulfill the Nyquist sampling criterion for the smallest scales probed. The SZ model image is then Fourier transformed and sampled to the position of the sparse interferometric data. The resulting synthetic visibilities are then employed in combination with the observed ones to evaluate the likelihood at each step of the Bayesian inference procedure.

However, we choose not to model the raw post-calibration data, instead binning the data in each spectral window following the optimal averaging scheme described in \citet{Hobson1995}. This is crucial for gaining a significant reduction in data volume and hence computational time.

\subsection{Sunyaev-Zeldovich model}\label{sec:sz}
A summary of the model priors introduced in this section can be found in Table~\ref{tab:tab01}. We test for biases in the parameter reconstruction arising from our specific choice for the distribution of priors by performing a \textit{prior-only} run, which is done by setting the likelihood to a constant value regardless of model fit \citep[see, e.g.,][]{DiMascolo2019}. As expected, the result of this test simply returns the input distribution of priors.

\begin{table}
    \centering
    \caption{Priors on the model parameters employed in our analysis.}
    \begin{tabular}{cclc}
        \hline\noalign{\smallskip}
        Param. & Prior & Details & Ref. \\\noalign{\smallskip}
        \hline\noalign{\medskip}
        \multicolumn{4}{l}{\textit{gNFW pressure model}}\\\noalign{\smallskip}
        $\mathrm{R.A.}$  & delta        & $\mu=\ra{6;58;35.6}$     &  1 \\\noalign{\smallskip}
        $\mathrm{Dec.}$ & delta        & $\mu=\dec{-55;57;10.8}$  &  1 \\\noalign{\smallskip}
        $P_{\mathrm{e,us}}$           & split-normal & $\mu=8.65\cdot10^{-3}~\mathrm{keV~cm^{-3}}$         &  2 \\
                                      &              &\hspace{-2pt}$\begin{aligned}\sigma=(&0.92\cdot10^{-3}~\mathrm{keV~cm^{-3}},\\[-2pt]&1.29\cdot10^{-3}~\mathrm{keV~cm^{-3}})\end{aligned}$ & \\\noalign{\smallskip}
        $T_{\mathrm{e,us}}$           & split-normal & $\mu=9.40~\mathrm{keV}$                             &  2 \\
                                      &              & \hspace{-1pt}$\sigma=(1.00~\mathrm{keV},~1.40~\mathrm{keV})$     &    \\\noalign{\medskip}\hline\noalign{\medskip}
        \multicolumn{4}{l}{\textit{Shock front}}\\
        \noalign{\smallskip}
        $\mathrm{R.A.}$         & split-normal & $\mu=\ra{6;58;15.5}$     &  3 \\
                                     &              & $\sigma=(\dec{;;2.3},~\dec{;;2.5})$ &    \\\noalign{\smallskip}
        $\mathrm{Dec.}$         & split-normal & $\mu=\dec{-55;56;58.26}$  &  3 \\
                                     &              & $\sigma=(\dec{;;8.6},~\dec{;;8.3})$ &    \\\noalign{\smallskip}
        $\theta$     & split-normal & $\mu=\dec{-98.14}$      &  3 \\
                                     &              & $\sigma=(\dec{4.52},~\dec{3.98})$ &    \\\noalign{\smallskip}
        $\mathcal{M}$                & uniform      & $\mathrm{min}=1$, $\mathrm{max}=10$                 & -- \\\noalign{\smallskip}
        $\alpha$                     & uniform      & $\mathrm{min}=-10$, $\mathrm{max}=0$                & -- \\\noalign{\medskip}\hline\noalign{\medskip}
        \multicolumn{4}{l}{\textit{Calibration}}\\\noalign{\smallskip}
        $\kappa_{\textsc{aca}}$      & normal       & $\mu = 1.00 $, $\sigma = 0.05$                      & -- \\\noalign{\smallskip}
        $\kappa_{\textsc{alma}}$     & normal       & $\mu = 1.00 $, $\sigma = 0.05$                      & -- \\\noalign{\medskip}
        \hline
    \end{tabular}
    \tablefoot{$\mu$ and $\sigma$ are the mode and the standard deviation of the probability distributions. The two values reported for $\sigma$ in the case of split-normal priors represent the standard deviations of the lower and upper halves of the corresponding distributions. The parameters $(\mathrm{R.A.},\mathrm{Dec.})_{\mathrm{gNFW}}$, $P_{\mathrm{e,us}}$, and $T_{\mathrm{e,us}}$ respectively define the centroid of the gNFW profile describing the upstream pressure distribution, and the upstream pressure, and the temperature normalization (see Sec.~\ref{sec:gnfw} for a discussion). For the shock front, we use its nose position $(\mathrm{R.A.},\mathrm{Dec.})_{\mathrm{shock}}$ as the reference point. Further, $\theta$ and $\mathcal{M}$ are the orientation of the shock axis and the Mach number (Sec.~\ref{sec:front}), while $\alpha$ is the slope of the downstream power-law profile (Sec.~\ref{sec:gnfw}). Finally, $\kappa_{\textsc{aca}}$ and $\kappa_{\textsc{alma}}$ are the ACA and ALMA calibration hyperparameters (Sec.~\ref{sec:gnfw}).}
    \tablebib{(1)~\citealt{Clowe2006}; (2)~X-ray model from \citealt{Markevitch2006} and \citealt{Markevitch2007}; (3)~This work (see Section~\ref{sec:sz}).}
    \label{tab:tab01}
\end{table}

\subsubsection{Shock front}\label{sec:front}
\begin{figure}
    \centering
    \includegraphics[trim= 3.1cm 0.0cm 2.4cm 0.0cm,clip,width=\columnwidth]{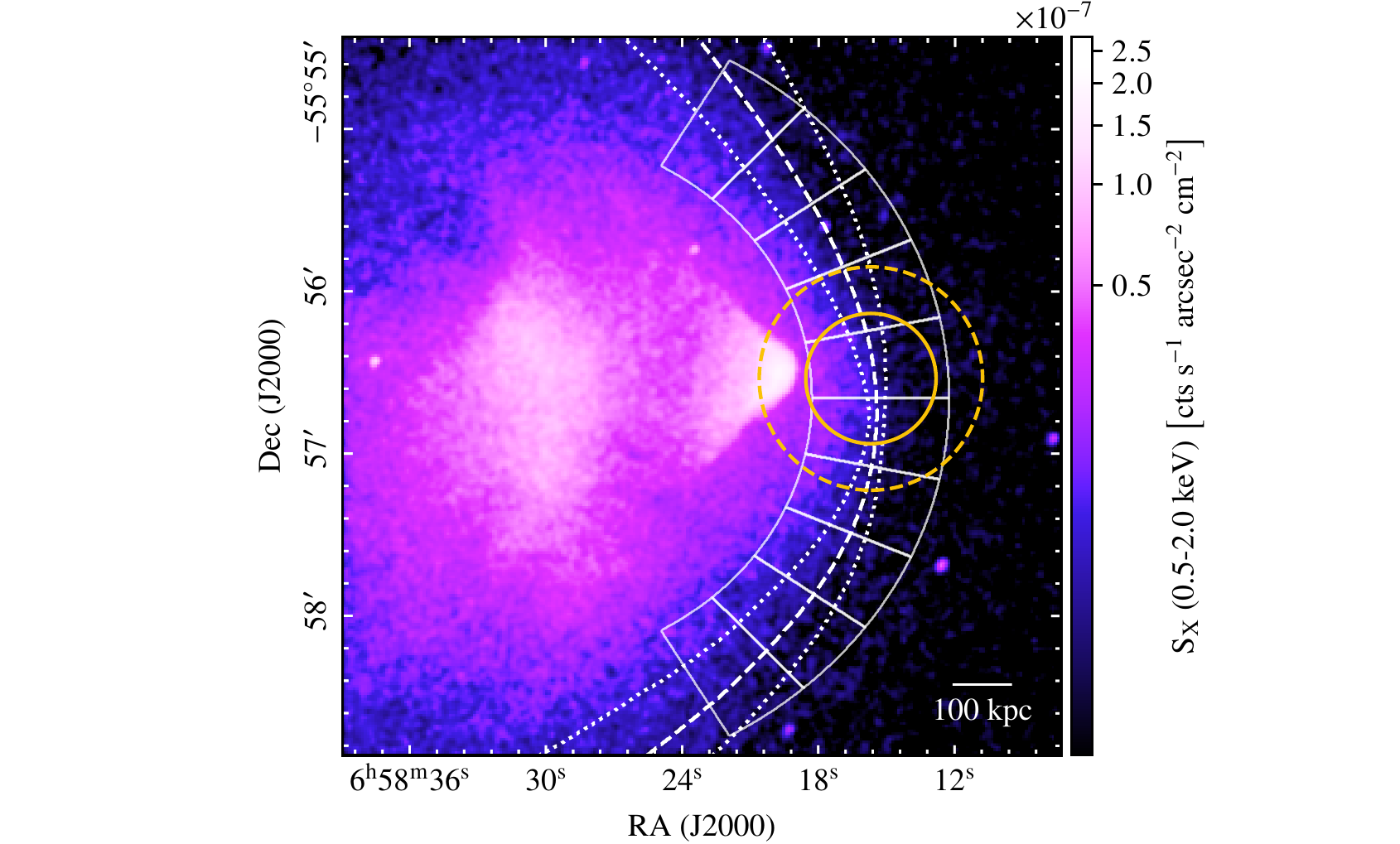}
    \caption{Cut-out of the $0.5-2.0~\mathrm{keV}$ \emph{Chandra} X-ray surface brightness map of the Bullet Cluster. The solid wedges represent the sectors employed to derive the hyperbolic shape best-matching the shock front geometry. The maximum-a-posteriori model is shown as a dashed line, while the dotted contours indicate the corresponding $95\%$ credible interval. Further, the dashed and solid yellow circles denote respectively the ACA and ALMA full-width-at-half-maximum fields of view. For reference, the upstream and downstream gas lie respectively west (right) and east (left) of the shock front.}
    \label{fig:figure02}
\end{figure}

{The common approach employed in the study of X-ray observations of shock fronts consists in describing them as spherical sectors within a specific region of the cluster image. This takes advantage of the image-space nature of the X-ray data to select a spatial region narrow enough to allow one to locally approximate the shock front as spherical. However, among the complexities of studying interferometric data is the difficulty of applying any spatial masking. This would entail convolution of the visibilities, inducing a non-trivial correlation between them. To avoid this, a complete two-dimensional model of the observed field is then required.}

In order to allow more freedom in the description of the shock front than in the case of a spherical model, we describe the shock front as an axially-symmetric hyperbolic surface (see the dashed line in Figure~\ref{fig:figure02}), with central axis coincident with the direction of the merger and lying in the plane of the sky. Since the interferometric data alone cannot constrain the line-of-sight distribution of pressure, we consider the curvature of the front to be symmetric with respect to the line-of-sight and the plane-of-sky direction. Although this is likely a reasonable assumption, any deviations from cylindrical symmetry may introduce non-negligible systematic errors into our results. In particular, the derived downstream pressure $P_{\mathrm{e,ds}}$ will be related to the true pressure $P\substack{\mathrm{true}\\\mathrm{e,ds}}$ as $P_{\mathrm{e,ds}}\approx P\substack{\mathrm{true}\\\mathrm{e,ds}} \left(\ell\substack{\mathrm{true}\\\mathrm{LoS}}/\ell_{\mathrm{LoS}} \right)$, where $\ell_{\mathrm{LoS}}$ and $\ell\substack{\mathrm{true}\\\mathrm{LoS}}$ are respectively the assumed and true line-of-sight extents of the shock front. We present specific estimates for the ratio $\ell\substack{\mathrm{true}\\\mathrm{LoS}}/\ell_{\mathrm{LoS}}$ when discussing our results in the next sections. It is worth noting that, as already discussed by \citet{Wang2018}, this is expected to scale approximately with the square root of the curvature radius of the shock surface. For line-of-sight and plane-of-sky curvatures radii $r_{\mathrm{LoS}}$ and $r_{\mathrm{PoS}}$, we would then get $\ell\substack{\mathrm{true}\\\mathrm{LoS}}/\ell_{\mathrm{LoS}} \approx \sqrt{r_{\mathrm{PoS}}/r_{\mathrm{LoS}}}$. We refer to Appendix~\ref{app:szscale} for a discussion.

A preliminary attempt to determine the shape of the shock front using SZ data alone shows that the parameters defining its geometry are heavily degenerate, and the small extent of the ALMA+ACA field of view does not allow for meaningful constraints. We would like to note this is a consequence of the sole parametrization of the shock geometry, since ALMA+ACA has proven to be able to identify edge positions with a beam-scale precision \citep{Basu2016}. We therefore derive a description of the morphology of the bow shock by finding the hyperbola that best describes the discontinuity observed in the \emph{Chandra} surface brightness map. Analogous to \citet{Ueda2017}, we find the best-matching shock geometry by minimizing the variance of the X-ray image within a defined region. The model is assumed to be simply given by a step function in which the discontinuity has a hyperbolic shape. The values of the function inside and outside the front itself are set equal to the mean photon counts in the respective regions of the X-ray image. In order to gain better leverage on the azimuthal geometry of the shock front, we further split the region in several angular sectors (see Fig.~\ref{fig:figure02}). The resulting maximum-a-posterior model for the hyperbolic surface is then employed for describing the profile of the shock front in all the following analyses. To account for a possible mismatch in the shock coordinates (e.g.\ due to astrometry errors) from the \emph{Chandra} modelling with respect to ALMA+ACA, we allow for some additional freedom in the nose coordinates $(\mathrm{R.A.},\mathrm{Dec.})_{\mathrm{shock}}$ and axis orientation $\theta$. Specifically, we assign each one priors based on the respective marginalized posteriors derived in the X-ray-matching step described above. Any asymmetry in the recovered parameter uncertainties is modelled by means of split-normal distributions \citep{Wallis2014}.

Apart from its morphology, the main parameter defining the shock is the ratio of downstream to upstream pressure at the jump itself. In practice, for a fixed line-of-sight geometry, the available ALMA+ACA data are mainly sensitive to the \textit{absolute difference} of the downstream and upstream electron pressures near the nose of the shock, i.e., $\Delta P_{\mathrm{e}}=P_\mathrm{{e,ds}}-P_{\mathrm{e,us}}$; the SZ signal associated with the large-scale distribution of the gas is effectively filtered out \citep[see, e.g.,][and discussion in Section~\ref{sec:collisional}]{Basu2016}. Thus, the modelling of the ALMA signal remains only weakly sensitive to the assumed large-scale model. The immediate downside is the \textit{relative} pressure jump at the shock, $x_{\textsc{p}}=P_{\mathrm{e,ds}}/P_{\mathrm{e,us}}=1+\Delta P_{\mathrm{e}}/P_{\mathrm{e,us}}$, which serves as a proxy for $\mathcal{M}$, is poorly constrained by the interferometric SZ data alone. In fact, due to the lack of information on the pressure normalization, the marginalized posterior distribution of the Mach number inferred when performing an SZ-only analysis are found to entirely span the corresponding prior interval. To get a meaningful measure of the pressure jump from the ALMA+ACA data, we therefore employ an X-ray-informed analysis of the ALMA+ACA SZ observations as in Section 3.3 of \citet{Basu2016}, and set the upstream electron pressure $P_{\mathrm{e,us}}$ to the value derived by modelling the \emph{Chandra} data in a narrow sector centered on the shock nose \citep{Markevitch2006}.

We do not include any model components describing the ``bullet'' itself (i.e., the contact discontinuity or ``cold front'') or the subtle additional cold front between the bullet and the main shock reported by \citet{Markevitch2006} and \citet{Markevitch2007}. The former lies outside the ALMA field of view, precluding any interesting constraints on the pressure difference (or lack thereof) across the cold front, while the latter is intrinsically faint, and is expected to be in thermal pressure equilibrium. For simplicity, we thus assume these features have a negligible effect on the measurements of the shock itself, and therefore ignore them. Further, we assume a single power-law profile for the downstream electron pressure (see below). A future analysis, joint with X-rays, will allow more model freedom for trying to build description of such features.

\subsubsection{Bulk pressure distribution}\label{sec:gnfw}
To model the pressure distribution in the downstream region, we employ a power law radial profile with slope $\alpha$, centered along the merger axis at a distance from the shock nose equal to the front curvature radius $r_{\mathrm{c}}$,
\begin{equation}
    P_\mathrm{e}(r) = x_{\textsc{p}} P_{\mathrm{e,us}} (r/r_{\mathrm{c}})^{\alpha}.
\end{equation}
On the other hand, we consider the pre-shock pressure distribution to be relaxed, thus to be described by a spherical generalized Navarro-Frenk-White (gNFW) profile \citep{Nagai2007}: 
\begin{equation}
    P_\mathrm{e}(r) \propto M\substack{a_{\textsc{p}}\\500} P_{500}\,P_{0}\,(c_{500} r/r_{500})^{-c}\,[1+(c_{500}r/r_{500})^{a}]^{(c-b)/a},
    \label{eq:gnfw}
\end{equation}
where $r_{500}$ and $P_{500}$ are functions of $M_{500}$, the total mass contained within an average overdensity 500$\times$ the critical density of the Universe at that redshift \citep{Arnaud2010}. We constrain $M_{500}$ so that the gNFW model always reproduces the X-ray value for the upstream pressure, $P_{\mathrm{e,us}}$. For the main results reported here, the pressure normalization $P_0$, concentration parameter $c_{500}$, mass-dependence index $a_{\textsc{p}}$, and slopes $(a,b,c)$ are fixed to the values reported in \citet{Arnaud2010} for the universal pressure profile. However, we show below that our results for $\mathcal{M}$ are insensitive to the choice of gNFW parametrization and position of the gNFW model centroid. We therefore simply fix the gNFW centroid's coordinates $(\mathrm{R.A.},\mathrm{Dec.})_{\mathrm{gNFW}}$ to the position of main lensing $\kappa$-map peak inferred by \citet{Clowe2006}, which we note does not coincide with the center of the post-shock profile.

\begin{figure}
    \centering
    \includegraphics[trim=0.4cm 0.4cm 0cm -0.120cm,clip,width=\columnwidth]{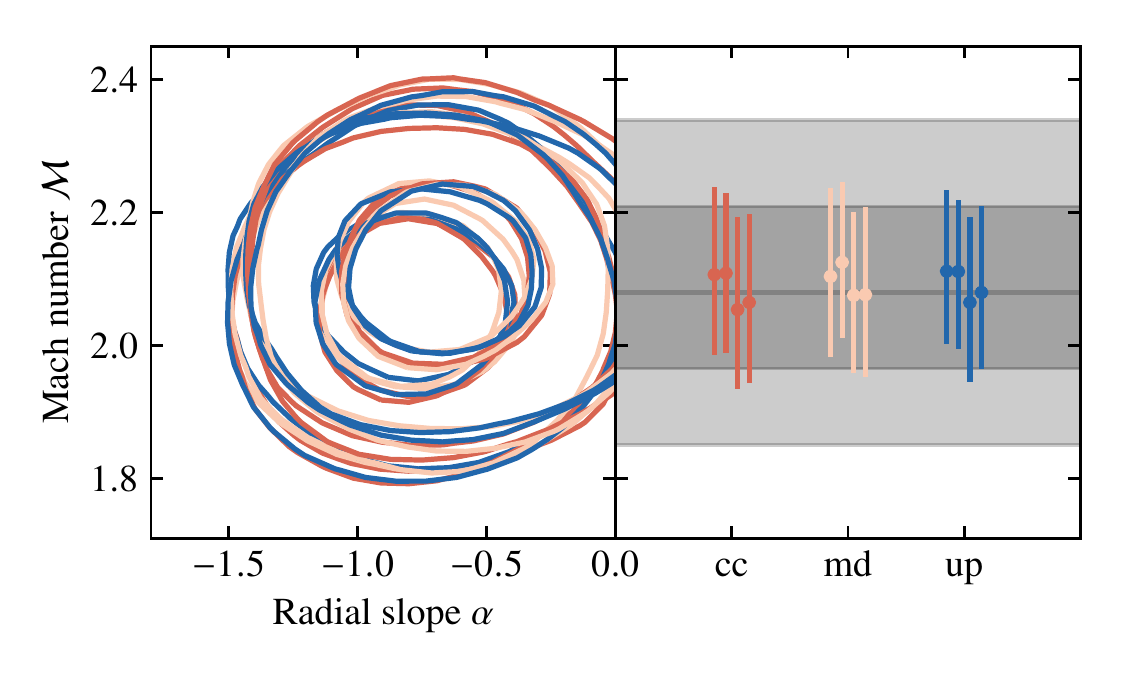}
    \caption{Bivariate posterior density function (left panel) for the inferred Mach number $\mathcal{M}$ and slope of the downstream pressure distribution $\alpha$ for a set of upstream pressure profiles and respective centroids. To facilitate comparison, the best-fitting $\mathcal{M}$ estimates are also plotted (right panel). We consider three different cases for the underlying gNFW profile by setting the slopes equal to the values reported in \citet{Arnaud2010} for the cool-core (cc; red), morphologically-disturbed (md; peach), or ensemble (up; blue) cluster samples. For each of them, the distribution centroid is then fixed to a number of different positions: far downstream and far upstream (right panel, left and mid-left points), respectively to arbitrary distances of $10\arcmin$ east and $3\arcmin$ west of the X-ray-derived shock nose coordinates; APEX-SZ centroid \citep[mid-right point;][]{Halverson2009}; peak of the $\kappa$-map \citep[right point;][]{Clowe2006}. In all the above cases, we assumed instantaneous shock heating of the electrons. The gray line in the right panel denotes the corresponding best-fitting $\mathcal{M}$ reported in Sec.~\ref{sec:instant}, while the darker and lighter bands the respective 68\% and 95\% credibility intervals.}
    \label{fig:figure01}
\end{figure}

Though pressure perturbations driven by the primary merger are confined to be within the shock front, it is possible for the passage of its associated dark matter component to affect, through infall, the ICM ahead of the shock \citep{Springel2007}. This may undermine our choice of the universal gNFW profile, reliable in the case of relaxed clusters, when describing the bulk pressure distribution. We tested against possible systematics introduced by this assumption. We found no significant deviations in the reconstructed parameters after changing either the slopes of the profiles or the position of the assumed centroids. The same applies to the structure of the gas on the downstream side as long as it is smooth, even though it may differ from expectations for a solid body moving through homogeneous fluid \citep[see, e.g.,][]{Zhang2019}. The results of the above tests are summarized in Fig.~\ref{fig:figure01}. The net result is that, as a consequence of the interferometric filtering, our shock model is largely sensitive to the pressure conditions right at the front, and not to the properties of the bulk pressure distribution (see Sec.~\ref{sec:collisional}). Thus, we consider wide uninformative priors on both the Mach number and post-shock slope, and marginalize over the latter.

In order to account for the high temperatures measured in the system, relativistic corrections to the SZ spectrum \citep{Itoh2004} are included in our modelling. In fact, variations in the measured SZ signal of the order of 5\% up to 15\% are expected for an electron gas with temperature ranging from $9~\mathrm{keV}$ to $30~\mathrm{keV}$ as measured from the X-ray data. As for $P_{\mathrm{e,us}}$, we employ an X-ray-motivated prior on the upstream temperature $T_{\mathrm{e,us}}$. In addition, we incorporate the $5\%$ uncertainties on the ACA and ALMA flux calibration by introducing normalization hyperparameters $\kappa_{\textsc{aca}}$ and $\kappa_{\textsc{alma}}$ \citep{DiMascolo2019}. For all the modelling runs presented in the following sections, $\kappa_{\textsc{aca}}$ and $\kappa_{\textsc{alma}}$ have been found to not deviate significantly from unity.

Given the plane-of-sky geometry of the merger involving the Bullet Cluster, any contribution from the kinetic SZ effect \citep{Sunyaev1980} to the observed signal due to the motion of the single subclusters should be subdominant with respect to the thermal SZ effect. In fact, if we assume the velocity $\varv\approx3000$ inferred from the shock Mach number \citep{Springel2007} to be measured with respect to the CMB rest frame and the merger direction to be oriented by around 8 degrees with respect to the plane of sky \citep{Markevitch2004}, we find that the contribution of the kinetic SZ effect to the total SZ signal from the post-shock region would be of the order of 3\% of the corresponding thermal component. This would induce a systematic error on the estimate of the Mach number $\mathcal{M}$ lower than 2\%. Given the small effect as well as the lack of robust constraints on the merger proper velocities and orientation, we then decide to not include the kinetic SZ effect in our model, keeping the merger axis aligned with plane of sky (see Sec.~\ref{sec:front}).

\subsection{Implementation details}
As noted earlier, the modelling algorithm and its specific implementation are detailed in \citet[][see also references therein]{DiMascolo2019}. However, we improved the posterior sampling algorithm by adopting the dynamic nested sampling by \citet{Higson2017}. In particular, we employ the pure-Python implementation provided by \texttt{dynesty} \citep{Speagle2019}.

\section{Results}
\begin{figure}
    \centering
    \includegraphics[width=0.97\columnwidth]{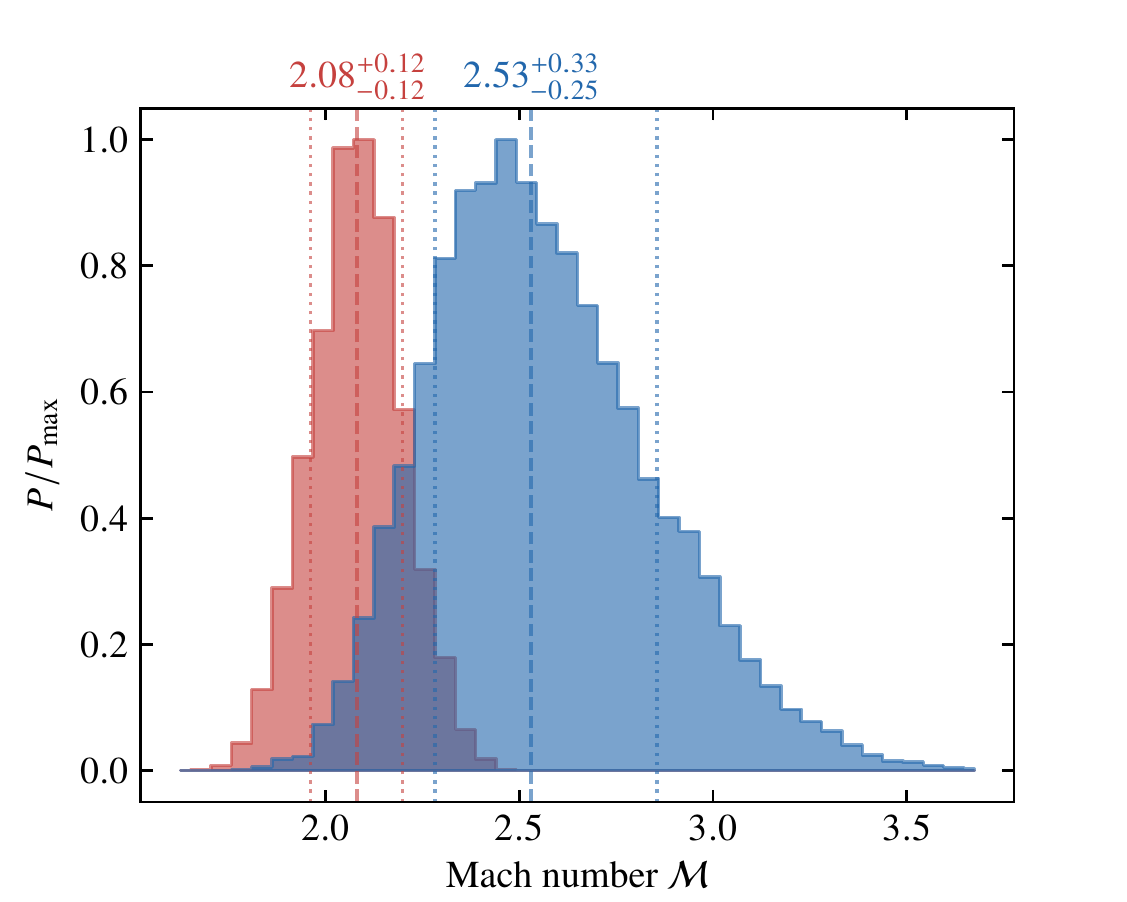}
    \caption{Marginalized posterior distributions for the shock Mach number $\mathcal{M}$ derived under the assumptions of instantaneous (red) and collisional (blue) electron-ion equilibration. The dashed and dotted lines indicates the median of the posterior distributions and the $68\%$ credible intervals, respectively.}
    \label{fig:figure03}
\end{figure}

\begin{figure*}
    \centering
    \includegraphics[width=\textwidth]{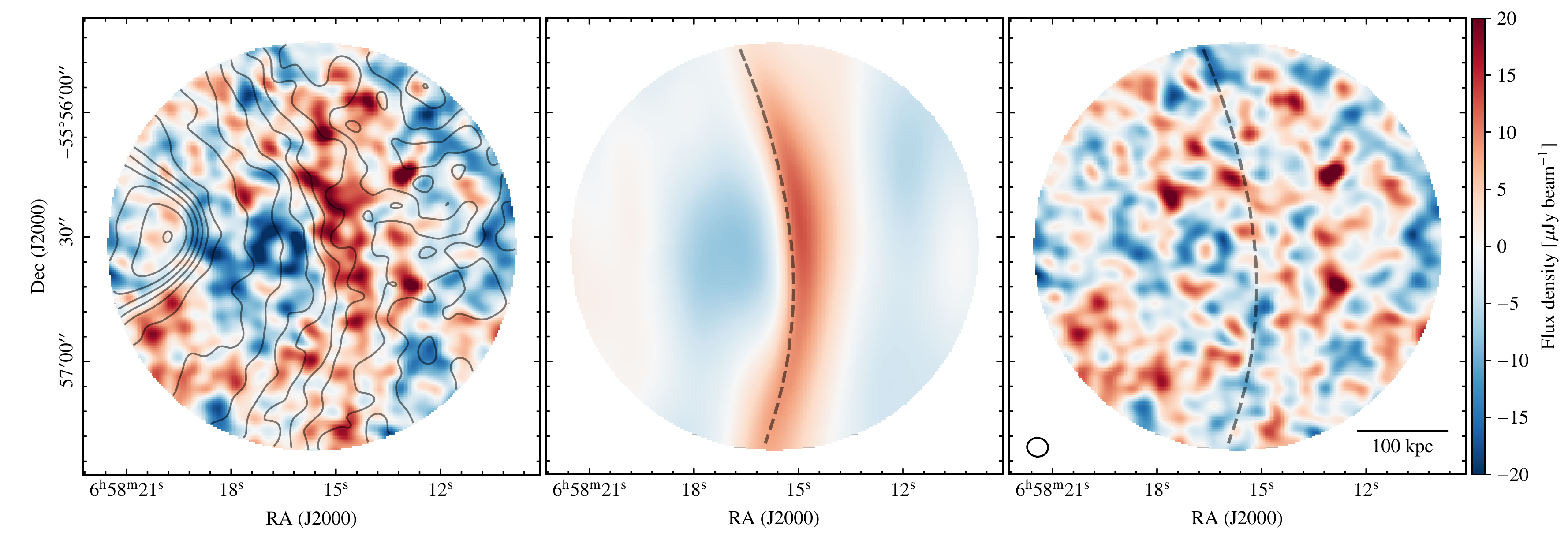}
    \caption{Dirty images of the raw (left), model (middle), and residual (right) ALMA+ACA interferometric data. They are generated by jointly gridding the ACA and ALMA data using a multi-frequency natural weighting scheme. For reference, this provides a synthesized beam of $4.07\arcsec\times3.01\arcsec$ FWHM (P.A.\ $81\degr$; bottom-left corner of right panel). We cut off the fields at the 0.2 gain level of the joint ALMA+ACA antenna pattern. To better highlight the large-scale shock features, we avoid correcting for the primary beam attenuation and apply an additional $30~\mathrm{k\lambda}$ taper.  We note that the model subtraction is performed directly in Fourier space. The dashed line in the center and right panels indicate the inferred position of the shock front. X-ray contours are overlaid on the left panel from Fig.~\ref{fig:figure02}.
    We also note the positive signal at the shock front is not due to an increment of the SZ signal; rather it is an artifact of the high-pass filtering effects of ALMA+ACA (see also Fig.~\ref{fig:figure05}).}
    \label{fig:figure04}
\end{figure*}

\subsection{Instantaneous electron-ion temperature equilibration}\label{sec:instant}
We first consider a model for the shock front under the standard assumption of an instantaneous electron-ion temperature equilibration, i.e., $T_e=T_i$, at the shock front. This is consistent with the \emph{Chandra} analysis by \citet{Markevitch2006}, who derived $\mathcal{M}_{\mathrm{X}}=3.0\pm0.4$ from the density jump, and measured an electron temperature jump as expected if electrons reach the average post-shock temperature near-instantly. A more recent analysis of the same \emph{Chandra} data by Markevitch (in prep.) provides an electron density jump $x_{\mathrm{n}}=2.86\pm0.16$, assuming instantaneous electron heating ($T_{\mathrm{e,ds}}\approx30~\mathrm{keV}$) for converting the X-ray surface brightness to the density jump by accounting for the different emissivity in the post- and pre-shock regions (see, e.g., \citealt{Ettori2000} for a discussion about X-ray brightness modelling in the presence of temperature gradients). This corresponds to $\mathcal{M}_{\mathrm{X}}=2.74\pm0.25$. The slight difference from the older estimate is due to a better-centered shock model and the inclusion of the azimuthal decline in $x_{\mathrm{n}}$ for angles away from the shock nose when projecting along the l.o.s. \citep[as in][]{Wang2018}, based on the amplitude of the density jumps measured from the X-ray data in different sectors of the shock (Markevitch, in prep.).

The assumption of instantaneous heating implies that the nominal Rankine-Hugoniot condition can be used to relate $\mathcal{M}$ to the measured amplitude of the electron pressure jump $x_{\textsc{p}}$ relative to the upstream value $P_{\mathrm{e,us}}$ as
\begin{equation}
  \mathcal{M} = \left[\frac{(\gamma+1)\,x_{\textsc{p}}+(\gamma-1)}{2\gamma}\right]^{1/2}.
  \label{eq:machRH}
\end{equation}
Here, $\gamma$ is the polytropic exponent, which we assume to be $\gamma=5/3$, appropriate for non-relativistic fully-ionized gas. We further allow for the azimuthal variation of $\mathcal{M}$ along the shock front, whose scaling with the azimuthal angle is derived using the same density jump decline discussed above. The omission of such azimuthal dependence would cause the Mach number to be averaged down with respect to its maximum value due to the effect of the wings with lower $\mathcal{M}$. For the results provided in this and the following sections, we estimate that the inclusion of the X-ray-based model for the azimuthal variation of the shock pressure jump increases the value of the inferred Mach number by only 5-7\%. However, more severe effects should be expected for observations with larger field of views, which would include values from farther in the wings.

We obtain $\mathcal{M}=2.08\substack{{+0.12}\\{-0.12}}$ (Fig.~\ref{fig:figure03}). While the model relies on the X-ray priors on the pre-shock pressure and temperature, the derived Mach number is inconsistent at a $2.4\sigma$ level with the X-ray estimate $\mathcal{M}_{\mathrm{X}}=2.74\pm0.25$. Projection effects may play a non-negligible role in biasing the SZ-based measurement of the Mach number. However, a strong ellipticity of the shock front shape  $\ell\substack{\mathrm{true}\\\mathrm{LoS}}/\ell_{\mathrm{LoS}} \lesssim 0.6$ (see Sec.~\ref{sec:front}) would be required to bridge the gap between SZ and X-ray estimates. In reality, an even larger ellipticity would be necessary, given that the X-ray estimates would also be affected by geometry, albeit with a different dependence. Another potential source of bias is the X-ray-motivated prior on the upstream pressure, which comes from deprojected density and temperature estimates, used to compute the relative pressure jump. While the definition of a centroid for X-ray deprojection remains ambiguous, we found only extreme choices would alter our results significantly. A joint-likelihood X-ray+SZ analysis may be required to find a consistent geometry that fully reconciles such discrepancies.

\subsection{Collisional electron-ion temperature equilibration}\label{sec:collisional}
Here we consider the possibility that the electron and ion temperatures do not equilibrate instantaneously in cluster shocks (i.e., $T_{\mathrm{e}} \neq T_{\mathrm{i}}$ immediately inside the shock front; see, e.g., \citealt{Fox1997}, \citealt{Markevitch2006}, \citealt{Russell2012}, \citealt{Wang2018}). Ions carry the majority of the gas bulk kinetic energy in collisionless shocks, and are heated dissipatively on scales comparable to their gyro-radii, while electrons might remain much colder \citep{Vink2015}, unless there is some process that equilibrates the ion and electron temperatures. The upper limit on the equilibration time scales is set by Coulomb collisions \citep{Zeldovich1966}, which for the downstream density and temperature in the Bullet Cluster is long ($\sim\mathrm{few}~10^8~\mathrm{yr}$), occurring over a distance comparable to the offset between the shock and the cold front.

Under the assumption of conservation of the enthalpy flux, electrons equilibrate with ions to the Rankine-Hugoniot downstream temperature at a rate driven by Coulomb collisions \citep{Fox1997}
\begin{equation}
    \frac{\mathrm{d}T_{\mathrm{e}}}{\mathrm{d}t} = \frac{1}{t_{\mathrm{eq}}} \left(1+\frac{n_{\mathrm{e}}}{n_{\mathrm{i}}}\right)\left(x_{\textsc{t}}T_{\mathrm{e,us}}-T_{\mathrm{e}}\right),
    \label{eq:eqrate}
\end{equation}
where $t_{\mathrm{eq}}$ is the Coulomb collisional time-scale \citep{Spitzer1962}, $n_{\mathrm{e}}$ and $n_{\mathrm{i}}$ are respectively the electron and ion densities, and $x_{\textsc{t}}$ is the temperature ratio across the shock front. To build our SZ model, we convert the above equation in terms of the distance from the shock front by means of the downstream gas velocity $u_{\mathrm{ds}}=(\mathcal{M}/x_{\mathrm{n}})\,c_{\mathrm{us}}$, with $c_{\mathrm{us}}$ given by the upstream sound speed. Furthermore, we assume that electrons are first heated adiabatically \citep{Vink2015}, so that the electron temperature immediately inside the shock front equals $x_{\mathrm{n}}^{\gamma-1} T_{\mathrm{e,us}}$. As required by the conservation of charge neutrality across the shock front, the density jump $x_{\mathrm{n}}$ is also set to follow the Rankine-Hugoniot condition. In this case, the pressure jump $x_{\textsc{p}}$ cannot be directly related to $\mathcal{M}$ as in Eq.~\ref{eq:machRH}, and instead must be derived as the product of the density and temperature ratios at each three-dimensional model coordinate.

In the case of collisional equilibration, we find $\mathcal{M}=2.53\substack{+0.33\\-0.25}$ (Fig.~\ref{fig:figure03}). This is consistent with $\mathcal{M}_{\mathrm{X}}=2.57\pm0.23$, coming from the \emph{Chandra} X-ray brightness fit if one uses the adiabatic-compression post-shock temperature ($T_{\mathrm{e,ds}}\approx 20~\mathrm{keV}$) to convert to the density jump (Markevitch, in prep.). Unfortunately, due to the severe filtering of large spatial scales as well as the limited field of view, we are not able to put any significant constraint on the specific equipartition time-scale when treating $t_{\mathrm{eq}}$ as a free parameter. Instead, we find that assuming the electron-ion equilibration to be driven by Coulomb collisions is practically equivalent to setting $t_{\mathrm{eq}}=\infty$.

\begin{figure}
    \centering
    \includegraphics[width=\columnwidth]{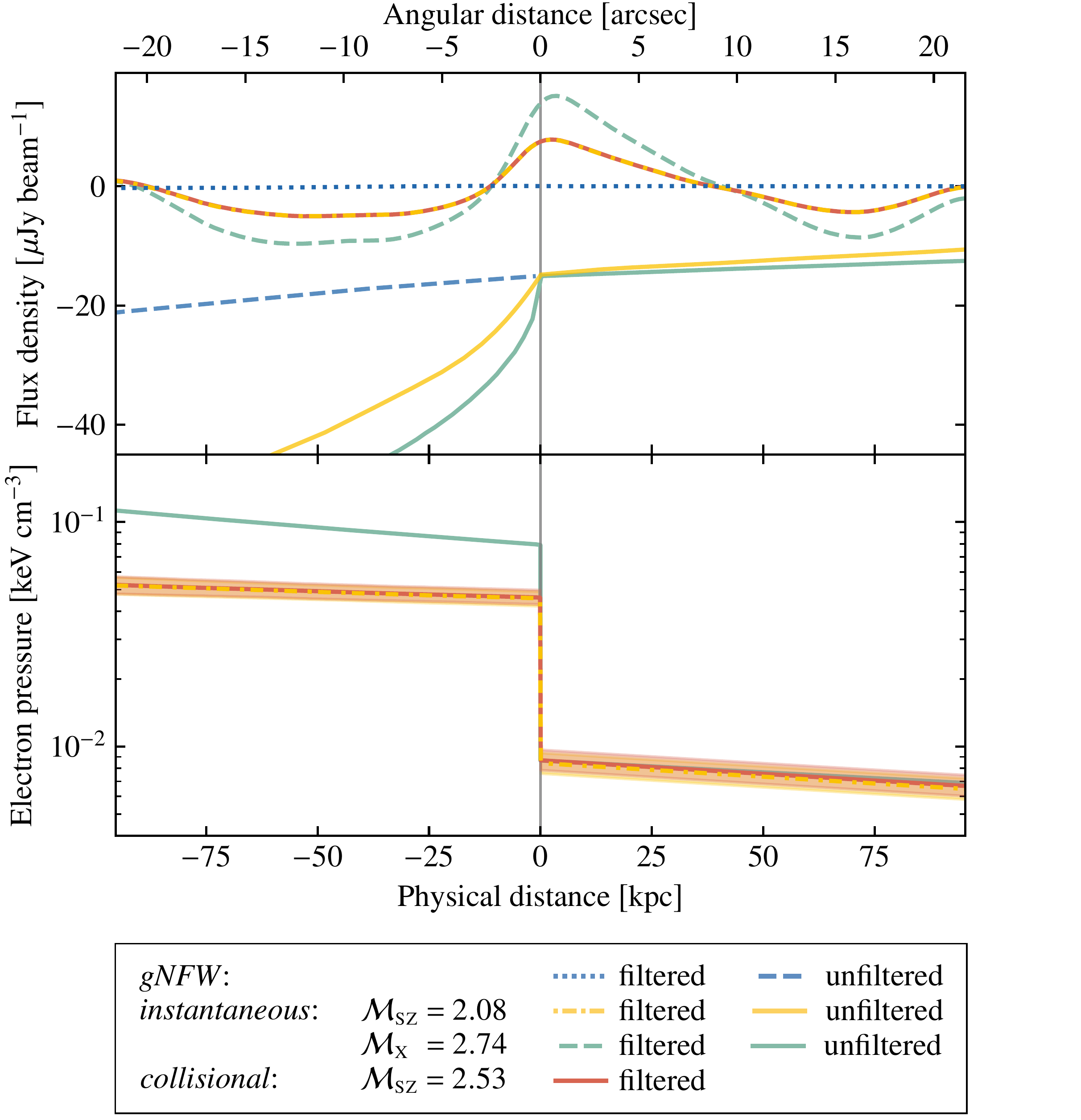}
    \caption{The figure provides a comparison of SZ signal profiles (upper panel) and the corresponding electron pressure profiles (lower panel) across the shock nose.
    The vertical gray line denotes the position of the shock front.
    We note that the upper panel contains both unfiltered input model fits for the SZ signal and the corresponding filtered (observed) profiles (see legend for details). As discussed in Sec.~\ref{sec:collisional}, the instantaneous (dash-dotted yellow) and collisional (solid red) shock models are indistinguishable after the spatial filtering of the interferometric ALMA+ACA observation, reflecting the fundamental limitation of ALMA+ACA to constrain any large-scale ($\gtrsim 1\arcmin$) component of the SZ signal. Again, we note that while we measure a decrement due to the SZ effect, the filtered (observed) profiles can exhibit both positive and negative excursions (analogous to the Gibbs phenomenon; see, e.g., \citealt{bracewell1978fourier}). For comparison, we also report the input unfiltered model (solid yellow) for the instantaneous equilibration, and both the raw and filtered underlying gNFW profile (dashed and dotted blue lines, as noted in the legend). We also present the X-ray expectation for both the filtered and raw SZ signal profiles (dashed and solid green lines) and corresponding pressure profiles that we would expect for the case of instantaneous equilibration, using the value $\mathcal{M}_{\mathrm{X}}=2.74\pm0.25$ derived from fits to the X-ray data.}
    \label{fig:figure05}
\end{figure}

For illustrative purposes, we present in Fig.~\ref{fig:figure04} the dirty\footnote{Here we refer to interferometric imaging closest to a raw Fourier transform of the visibilities.  For these images, no attempt to apply the \textsc{clean} algorithm to deconvolve the image of the synthesized beam has been made.} images of the raw ALMA+ACA data employed in our analysis, the interferometric model corresponding to the collisional electron-ion equilibration scenario, and the respective model-subtracted data. As shown in the right panel, it is not possible to identify residuals that differ at a significant level from noise-like features. 

The analogous image for the instantaneous case is visually identical to Fig.~\ref{fig:figure04}, and therefore is not shown. This is evident in Fig.~\ref{fig:figure05}, where it is not possible to identify any significant difference between the filtered SZ models for the instantaneous- and collisional-equilibration scenarios. This confirms that the ALMA+ACA data are only sensitive to the properties of the pressure distributions near the shock edge, thus providing a direct estimate of the pressure difference across the shock front, i.e., $\Delta P_{\mathrm{e}}$ rather than  $x_{\textsc{p}}$. Moreover, our result is found to be practically independent of the specific assumption about the underlying gNFW profile, which is entirely filtered by the interferometric response (Fig.~\ref{fig:figure05}). 

The fact ALMA+ACA is only sensitive in practice to the electron pressure difference $\Delta P_{\mathrm{e}}$ across the shock is also reflected in the lack of any significant difference between the Bayesian evidences of the instant equilibration and adiabatic heating models ($\Delta\log{\mathcal{Z}}=1.30\substack{+0.71\\-1.48}$). In fact, given that we cannot observe any large-scale feature in the SZ signal induced by the slow increase of the post-shock electron temperature in the case of collisional equilibration, the different heating scenarios practically differ only in the way we convert the pressure difference to an estimate of the shock $\mathcal{M}$.

We recall here a subtle cold front is observed in the \emph{Chandra} X-ray image between the shock and the bullet boundary, $\sim15''$ east from the shock. Since the total pressure across a cold front is expected to be approximately continuous, this was not included in our modeling of the downstream pressure profile. In fact, no apparent signature of such feature can be distinguished in the ALMA+ACA SZ observation. However, because the flow of the post-shock gas would not cross the cold front, it is unlikely that any electron-ion temperature non-equilibrium would extend past it. If indeed there is an electron-proton temperature difference in the post-shock region, we should expect the electron temperature (and, hence, pressure) to reach its equilibrium value. This would in turn result in a discontinuity in the SZ signal in the direction of the front itself. Future, more sensitive ALMA observations may search for such a feature.

\section{Conclusions}
We further demonstrate the ability of using deep, high-resolution ALMA+ACA observations of the SZ effect to characterize shocks in merging clusters \citep[see, for comparison,][]{Basu2016}. For this purpose, we studied the SZ effect across the shock in the Bullet Cluster, chosen as it is widely regarded as the ``textbook example'' of a cluster merger bow shock.

The application of our interferometric modelling technique -- using X-ray-motivated priors -- has allowed us to place constraints on the electron pressure discontinuity across the shock.  Assuming a Rankine-Hugoniot shock adiabat, our pressure jump implies a Mach number $\mathcal{M}=2.08\substack{+0.12\\-0.12}$, which is significantly lower than the one derived from \emph{Chandra} data using the same geometric assumptions ($\mathcal{M}=2.74\pm0.25$). An interesting physical possibility to reconcile the two measurements is to allow that the electron and ion temperatures do not equilibrate instantly after the shock passage has heated the electrons adiabatically. For a given Mach number, this would lower the post-shock electron temperature and thus the observed electron pressure jump. Our Mach number would then become $\mathcal{M}=2.53\substack{+0.33\\-0.25}$, in agreement with the X-ray estimate that assumes the adiabatic temperature jump for conversion between the X-ray brightness and density. We note that \emph{Chandra} X-ray data constrain the gas density (from which the Mach number is derived) and electron temperature across the shock separately, and its post-shock temperature prefers instant equilibration over adiabatic heating of the electrons (at $\sim 2\sigma$ confidence; \citealt{Markevitch2006}). However, while \emph{Chandra} is free from ALMA+ACA interferometric limitations and can probe the upstream and downstream gas directly, the Bullet post-shock temperature is above the range where \emph{Chandra} can measure electron temperatures reliably, and hence suffer significant systematic uncertainties.

To summarize, ALMA+ACA has proven to provide a clean measurement of the differential jump in pressure due to the shock, and, in combination with data that can access larger scales, can provide compelling constraints on shock properties such as the Mach number. In particular:
\begin{itemize}
  \item Interferometric observations cleanly measure the projected pressure jump due to the shock. However, due to the inherent spatial filtering of ALMA and the ACA, which recover scales $\sim0.5-1.1\arcmin$ in Band 3 (compared to $\theta_{500}\sim $10\arcmin, corresponding to $r_{500}$ for the Bullet Cluster), X-ray priors on both the model geometry and upstream pressure are necessary in order to infer $\mathcal{M}$. On the other hand, the combination of SZ observations covering a broader range of angular scales (i.e., from $0.1\arcsec-30\arcmin$) could provide an SZ-only view of the shock properties.\\
  \item Once the geometry is fixed, the key quantities which drive the analysis and the interpretations of the results are the pressure difference across the shock front, the normalization of the pre-shock pressure, and the independent X-ray estimates of $\mathcal{M}$. We show their statistical uncertainties are small enough to allow us to differentiate between the instantaneous and adiabatic heating scenarios. Nevertheless, neither model is unambiguously preferred. Although the two scenarios result in SZ-based estimates for $\mathcal{M}$ that deviate one from the other by $\sim2\sigma$, we find that the difference of the respective Bayesian log-evidence is not significant enough to completely rule out one versus another.\\
  \item We extensively tested our modelling choices --- varying the geometry, pre- and post-shock pressure slopes, and underlying pressure distribution --- and find our results to be robust for a broad range of possible assumptions motivated by the X-ray analyses. However, our model does not fully describe the complex morphology observed in the X-ray surface brightness. Together with the uncertainties on the three-dimensional morphology of the cluster, this may limit our ability to elucidate the nature of electron heating across the shock front.
\end{itemize}

Together, these illustrate the fundamental complementarity of X-ray and SZ effect observations in the study of the physics of galaxy clusters. It is then clear that a simultaneous, joint-likelihood analysis of the SZ and X-ray data on the Bullet Cluster, extending the approach of incorporating X-ray information in the form of priors \citep[see our discussion in Sec.~\ref{sec:front}, and Section 3.3 of][]{Basu2016}, would benefit our understanding of the morphology of the galaxy cluster, as well as provide further insights into the physical mechanisms for shock heating of the intracluster medium. A forthcoming paper will present the results of a full joint-likelihood analysis of interferometric SZ and X-ray observations, as well as single-dish SZ measurements, building on the methodology discussed in \citet{DiMascolo2019}. Meanwhile, upcoming results from NuSTAR (Wik et al. in prep) will better access the high photon energies corresponding to the high temperatures inferred from \emph{Chandra}. Further, the number of observations of shocks with unambiguous geometry and sufficiently high Mach number that allow the detection of deviations from instantaneous electron heating is limited. Thus, along with improved modelling, observations of a larger sample of cluster shocks will be needed to improve our understanding. And finally, both improved spatial and spectral resolution, larger instantaneous field of view, and the ability to recover zero-spacing information will vastly improve future SZ-only studies. However, in order to provide sufficient overlap with the interferometric data in Fourier space, while also probing higher frequencies and angular scales $>$10\arcmin, a new wide-field ($> 1^\circ$) single-dish facility, such as the Atacama Large Aperture Submm/mm Telescope (\href{http://atlast-telescope.org}{AtLAST}; see, e.g., \citealt{Klaassen2019,Mroczkowski2019b}) is required.

\section*{Acknowledgements}
We thank the anonymous referee for the insightful comments.

EC and RS acknowledges partial support from the Russian Science Foundation grant 19-12-00369.
This paper makes use of the following ALMA data: ADS/JAO.ALMA\#2013.1.00760.S. ALMA is a partnership of ESO (representing its member states), NSF (USA) and NINS (Japan), together with NRC (Canada), MOST and ASIAA (Taiwan), and KASI (Republic of Korea), in  cooperation with the Republic of Chile. The Joint ALMA Observatory is operated by ESO, AUI/NRAO and NAOJ. Basic research in radio astronomy at the US Naval Research Laboratory is funded by 6.1 Base funding, supporting T. Clarke.

The scientific results reported in this article are obtained using the following softwares: \texttt{AstroPy}, a community-developed core Python package for Astronomy \citep{Astropy2018}; \texttt{dynesty} \citep{Speagle2019}; \texttt{galario} \citep{Tazzari2018}.

\bibliographystyle{aa} 
\bibliography{bullet.bib}

\appendix
\section{Interferometric view of the SZ effect}\label{app:szscale}
The amplitude of the thermal SZ effect in the direction $\bm{x}$ of a galaxy cluster characterized by an electron pressure distribution $P_{\mathrm{e}}$ is proportional to Compton $\vary$,
\begin{equation}
    \vary(\bm{x}) = \frac{\sigma_{\textsc{t}}}{m_{\mathrm{e}}c^2} \int P_{\mathrm{e}}(\bm{x},\ell)\, \mathrm{d}\ell,
    \label{eq:compton}
\end{equation}
where $\sigma_{\textsc{t}}$, $m_{\mathrm{e}}$, and $c$ are respectively the Thomson cross-section, the electron mass, and the speed of light, while $\ell$ is the coordinate along the line of sight. 

As shown in Sec.~\ref{sec:collisional} and Fig.~\ref{fig:figure05}, the available ALMA+ACA data (i) only probe a small region near the tip of the shock and (ii) any extended structures in this region are effectively filtered out. Therefore, a model that includes a jump of pressure across the shock front and is smooth otherwise should capture the most of the information contained in the ALMA+ACA data. To this end, we represent the electron pressure distribution $P_{\mathrm{e}}(\bm{x},\ell)$ as a combination of two spatially smooth components, $P_{\mathrm{e,0}}(\bm{x},\ell)$ and $P_{\mathrm{e,1}}(\bm{x},\ell)$, where their sum is
\begin{equation}
    P_{\mathrm{e}} = (1-f_{\mathrm{ds}})\, P_{\mathrm{e,0}} +f_{\mathrm{ds}}\, P_{\mathrm{e,1}}.
    \label{eq:comptony}
\end{equation}
Here $f_{\mathrm{ds}}\equiv f_{\mathrm{ds}}(\bm{x},\ell)$ is equal to $1$ in the downstream region and $0$ in the upstream region (i.e., is a Heaviside step function). We note both that here and below, for simplicity, we omit the explicit coordinate dependence of $P_{e}$ or $y$. Rearranging the terms and integrating along the line of sight we obtain
\begin{equation}
    y \propto {\textstyle\int P_{\mathrm{e,0}} \, \mathrm{d}\ell} + {\textstyle\int (P_{\mathrm{e,1}}-P_{\mathrm{e,0}})\, f_{\mathrm{ds}}\, \mathrm{d}\ell}.
    \label{eq:comptony2}
\end{equation}
Since the first term in the above expression corresponds to a smooth, large-scale pressure distribution 
its contribution to $y$ is filtered out from the ALMA+ACA data. 
Moreover, the function $(P_{\mathrm{e,1}}-P_{\mathrm{e,0}})$ in the second term is also spatially smooth and would be filtered too without the step function $f_{\mathrm{ds}}$.  Therefore, the signal $\tilde\vary$ measured by ALMA+ACA in the vicinity of the shock tip is effectively defined by the second term in equation~\ref{eq:comptony2}, which is set by the pressure jump at the shock front and the length-scale of the downstream region.  Thus,
\begin{equation}
    \tilde\vary \propto  {\textstyle\int (P_{\mathrm{e,1}}-P_{\mathrm{e,0}})\, f_{\mathrm{ds}}\, \mathrm{d}\ell} \approx (P_{\mathrm{e,ds}}-P_{\mathrm{e,us}})\,\ell_{\mathrm{LoS}}=\Delta P_{\mathrm{e}} \, \ell_{\mathrm{LoS}},
    \label{eq:comptonytilde}
\end{equation}
where $\ell_{\mathrm{LoS}}$ is the line-of-sight extent of the probed post-shock region, and $P_{\mathrm{e,us}}$ and $P_{\mathrm{e,ds}}$ are the electron pressures measured just outside and inside the shock front, respectively (Sec.~\ref{sec:sz}). Therefore, ALMA+ACA data effectively constrain a product of the electron pressure difference at the shock $\Delta P_{\mathrm{e}}$ and the physical size of the region $\ell_{\mathrm{LoS}}$. The latter quantity can be easily determined if the merger is in the plane of the sky and the shock front possesses rotational symmetry. If the shape of the front can be approximated by a sphere with a radius $R$, then along the symmetry axis $\ell_{\mathrm{LoS}}\approx \sqrt{2 r h}$, where $h \ll r$ is the distance from the tip of the shock. While the calculations in the paper were done without these simplifying assumptions, the equation~\ref{eq:comptonytilde} is useful to estimate the uncertainty introduced by the (unknown) geometry of the shock along the line of sight. In particular, if the curvatures in the sky plane $r_{\mathrm{PoS}}$ and along the line of sight $r_{\mathrm{LoS}}$ differ, the estimate of the pressure difference $\Delta P_{\mathrm{e}}$, which assumes $r_{\mathrm{LoS}}=r_{\mathrm{Pos}}$, will be biased by a factor $\sqrt{r_{\mathrm{LoS}}/r_{\mathrm{PoS}}}$ (the same argument is discussed in \citealt{Wang2018}).

\end{document}